\documentstyle[eqsecnum,preprint,aps,amsfonts]{revtex}
\begin{document}

\draft

{\tighten
\preprint{\vbox{\hbox{CALT-68-1996}
                \hbox{hep-ph/9506201}
		\hbox{\footnotesize DOE RESEARCH AND}
		\hbox{\footnotesize DEVELOPMENT REPORT} }}

\title{Moments of the photon spectrum in the \\
  inclusive $B\to X_s\,\gamma$ decay\footnote{%
Work supported in part by the U.S.\ Dept.\ of Energy under Grant no.\
DE-FG03-92-ER~40701.} }

\author{Anton Kapustin and Zoltan Ligeti}

\address{California Institute of Technology, Pasadena, CA 91125}

\maketitle

\begin{abstract}%
The moments of the photon spectrum in the inclusive $B\to X_s\,\gamma$ decay
can be calculated to order $\alpha_s$ accuracy at present, without knowing the
$\alpha_s$ corrections to the effective Hamiltonian.  We discuss the standard
model predictions, and how the moments of the photon spectrum are related to
certain matrix elements of the heavy quark effective theory.  The sensitivity
of these moments to new physics is small, and they provide a model-independent
determination of the $b$ quark pole mass (at order $\alpha_s$), or
equivalently, the matrix elements $\bar\Lambda$ and $\lambda_1$ of the
effective theory.
\end{abstract}

}

\newpage

\section{Introduction}

The inclusive $B\to X_s\,\gamma$ decay has received a lot of attention in
recent years \cite{old,GSW,LLOb,NLO,BMMP,YY}, primarily due to its sensitivity
to physics beyond the standard model (SM) \cite{GSW,BMMP,YY}.  As any flavor
changing neutral current process, it can only arise at one-loop level in the
SM, and therefore possible new physics can yield comparable contributions.
However, the recent CLEO measurement \cite{CLEO} excludes large deviations from
the SM.

Since the $b$ quark is heavy compared to the QCD scale, the inclusive $B\to
X_s\,\gamma$ decay rate can be calculated in a systematic QCD-based expansion
\cite{CGG}.  The decay rate computed in the $m_b\to\infty$ limit coincides with
the free quark decay result.  Corrections can then be included in an expansion
in powers of $1/m_b$ and $\alpha_s(m_b)$.

At present, the theoretical prediction for the decay rate suffers from large
uncertainties, as the result is only known in the leading logarithmic
approximation.  To refine the theoretical prediction and thus increase the
sensitivity to new physics, a next-to-leading order calculation is needed.
That is a very demanding task, as it requires the evaluation of many two-loop
and even the infinite parts of three-loop diagrams.  In the absence of such a
calculation, and since the recent CLEO result shows no evidence for new physics
contributing significantly to the $B\to X_s\,\gamma$ decay, we investigate what
the presently available measurement could teach us.

We point out that the moments of the photon spectrum can be obtained to order
$\alpha_s$ accuracy by a relatively simple calculation.  We evaluate these
corrections to the first few moments of the photon spectrum.
Since $B\to X_s\,\gamma$ is a two-body decay at the quark level (at leading
order in $\alpha_s$), the photon spectrum is monochromatic in the spectator
model.  Therefore, the moments are also sensitive to the nonperturbative
corrections in the heavy quark expansion.  They provide a model-independent
determination of the $b$ quark pole mass, {\it i.e.}, the matrix elements
$\bar\Lambda$ and $\lambda_1$ of the heavy quark effective theory (HQET).

\section{The effective Hamiltonian}

The $B\to X_s\,\gamma$ decay in the standard model is mediated by penguin
diagrams.  The QCD corrections to this process form a power series in the
parameter $\alpha_s\ln(M_W^2/m_b^2)$, that is too large to provide a reliable
expansion.  Therefore, it is convenient to integrate out the virtual top quark
and $W$ boson effects (and possible new physics) at the $W$ scale, and sum up
the large logarithms using the operator product expansion and the
renormalization group.  We work with the operator basis and effective
Hamiltonian of Ref.~\cite{GSW}
\begin{equation}
H_{\rm eff} = -{4G_F\over\sqrt2}\, V_{ts}^*\,V_{tb}\,
  \sum_{i=1}^8 C_i(\mu)\, O_i(\mu) \,,
\end{equation}
where
\begin{equation}\label{ops}
  \begin{array}{ll}
O_1 = (\bar c_{L\beta}\,\gamma^\mu\, b_{L\alpha})\,
  (\bar s_{L\alpha}\,\gamma_\mu\, c_{L\beta})  \,, &\qquad
O_2 = (\bar c_{L\alpha}\,\gamma^\mu\, b_{L\alpha})\,
  (\bar s_{L\beta}\,\gamma_\mu\, c_{L\beta})  \,, \\
\displaystyle
O_7 = {e\over16\pi^2}\,m_b\, \bar s_{L\alpha}\,\sigma^{\mu\nu}\,F_{\mu\nu}\,
  b_{R\alpha} \,, &\qquad
\displaystyle
O_8 = {g\over16\pi^2}\,m_b\, \bar s_{L\alpha}\, \sigma^{\mu\nu}\,
  T_{\alpha\beta}^a\, G_{\mu\nu}^a\, b_{R\beta} \,.
  \end{array}
\end{equation}
We listed here only the operators whose Wilson coefficients are of order unity;
$C_3-C_6$ are about an order of magnitude smaller, since they arise only due to
operator mixing.  As the matrix elements of all operators except $O_7$ contain
an overall factor of $\alpha_s$ (once we use the ``effective" Wilson
coefficients \cite{BMMP} defined below), we shall neglect $O_3-O_6$.

The order $\alpha_s$ corrections to the the photon spectrum
come from three sources \cite{BMMP}:
\vspace{-8pt}
\begin{itemize}\itemsep=-4pt
\item[(i)]
Corrections to $C_i(M_W)$ coming from the matching of the SM matrix elements
onto the effective Hamiltonian: the order $\alpha_s$ corrections to $C_7(M_W)$
and $C_8(M_W)$ are not yet known.
\item[(ii)]
Corrections to the running of the Wilson coefficients $C_i$ between the $W$
and the $m_b$ scale: the next-to-leading order mixing of the dimension six with
the dimension five operators resulting from three-loop diagrams is not known at
present.
\item[(iii)]
The $\alpha_s$ corrections to the matrix elements of the operators in the
effective Hamiltonian at the low scale: these corrections also involve
unknown two-loop diagrams.
\end{itemize}
\vspace{-8pt}
Certain parts of the next-to-leading log anomalous dimension matrix (ii) have
been calculated \cite{NLO}.  The unknown contributions are expected to be
significant, as these terms should reduce the large $\mu$-dependence of the
leading log result.  The last class of corrections (iii) has been considered in
Ref.~\cite{AG}.  However, important two-loop diagrams involving $O_2$, that
contribute to the spectrum near the maximal photon energy, were neglected.
Moreover, in the absence of the next-to-leading order result for $C_7(\mu)$,
any calculation of the spectrum inevitably contains renormalization scheme
dependence at order $\alpha_s$, and is therefore ill-defined.\footnote{We
thank Mark Wise for pointing this out to us.}

Since none of these three ingredients of the full order $\alpha_s$ calculation
are known, it may be surprising that the moments can be calculated to this
accuracy.  We prove this in the next section.  Thus, for our purposes it is
sufficient to summarize the results for the coefficients of the effective
Hamiltonian in the leading logarithmic approximation, for which we adopt the
scheme independent definitions of\cite{BMMP}.  In the standard model
$C_2(M_W)=1$, and
\begin{eqnarray}
C_7(M_W) &=& {3x^3-2x^2\over4(x-1)^4}\,\ln x + {-8x^3-5x^2+7x\over24(x-1)^3}
  \,,\nonumber\\
C_8(M_W) &=& {-3x^2\over4(x-1)^4}\,\ln x + {-x^3+5x^2+2x\over8(x-1)^3} \,,
\end{eqnarray}
where $x=m_t^2/M_W^2$.  At a low scale $\mu$, these coefficients become
\begin{eqnarray}\label{coeffs}
C_2(\mu) &=& \frac12\, \Big(\eta^{6/23} + \eta^{-12/23}\Big)\,
  C_2(M_W) \,,\nonumber\\
C_7^{\,{\rm eff}}(\mu) &=& \eta^{16/23}\,C_7(M_W)
  + \frac83\, \Big(\eta^{14/23}-\eta^{16/23}\Big)\, C_8(M_W)
  + C_2(M_W)\, \sum_{i=1}^8 h_i\,\eta^{a_i} \,,\nonumber\\
C_8^{\,{\rm eff}}(\mu) &=& \eta^{14/23}\,C_8(M_W)
  + C_2(M_W)\, \sum_{i=1}^5 g_i\,\eta^{b_i} \,,
\end{eqnarray}
where $\eta=\alpha_s(M_W)/\alpha_s(\mu)$, and the numerical values of the
$h_i$'s, $g_i$'s, $a_i$'s, and $b_i$'s can be found in \cite{BMMP}.
The scale is usually chosen to be $\mu=m_b$, and one estimates the
uncertainties related to the unknown higher order terms by varying $\mu$,
typically between $m_b/2$ and $2m_b$.  However, our results will be
scale independent to order $\alpha_s$, so this is not going to be a large
uncertainty.

\section{Moments}

In this section we show that moments of the photon spectrum can be calculated
to order $\alpha_s$, although none of the three previously described
ingredients of the full order $\alpha_s$ computation of the decay rate are
known.  Since experimentally one needs to make a lower cut on the photon
energy, we define the moments of the photon spectrum as
\begin{equation}\label{Moments}
M_n(E_0) = {\displaystyle \int_{E_0}^{E_\gamma^{\rm max}}\! E_\gamma^n\,
  {{\rm d}\Gamma\over{\rm d}E_\gamma}\, {\rm d}E_\gamma \over \displaystyle
  \int_{E_0}^{E_\gamma^{\rm max}}\! {{\rm d}\Gamma\over{\rm d}E_\gamma}\,
  {\rm d}E_\gamma } \,.
\end{equation}
Here $E_\gamma^{\rm max}=[M_B^2-(M_{K}+M_\pi)^2]/2M_B$ is the maximal possible
photon energy.  To illustrate our argument, we denote schematically the
contribution of a given operator $O_i$ to the $n$-th moment of the spectrum
by $\langle O_i\rangle_n$.  Then we can rewrite the moments $M_n$ as
\begin{equation}
M_n \sim \left| {C_7^{\,{\rm eff}}\,\langle O_7\rangle_n +
  C_8^{\,{\rm eff}}\,\langle O_8\rangle_n +
  C_2\,\langle O_2\rangle_n \over C_7^{\,{\rm eff}}\,\langle O_7\rangle_0 +
  C_8^{\,{\rm eff}}\,\langle O_8\rangle_0 + C_2\,\langle O_2\rangle_0}\right|^2
= \left| {\displaystyle \langle O_7\rangle_n + {C_8^{\,{\rm eff}} \over
  C_7^{\,{\rm eff}}}\,\langle O_8\rangle_n +
  {C_2\over C_7^{\,{\rm eff}}}\,\langle O_2\rangle_n \over \displaystyle
  \langle O_7\rangle_0 + {C_8^{\,{\rm eff}}\over C_7^{\,{\rm eff}}}\,
  \langle O_8\rangle_0 + {C_2\over C_7^{\,{\rm eff}}}\,\langle O_2\rangle_0}
  \right|^2 \,.
\end{equation}
For $i\neq7$, the contributions $\langle O_i\rangle_n$ are of order $\alpha_s$.
Therefore, to determine the order $\alpha_s$ corrections to $M_n$, it is
consistent to take into account these contributions, and the $\alpha_s$
correction to the matrix element of the operator $O_7$ as well.  But it is
sufficient to know the Wilson coefficients $C_i$ only to the presently
available (leading log) accuracy.

The other important observation is that the two-loop part of the order
$\alpha_s$ contributions to matrix elements of the operators at the low energy
scale do not contribute to the moments of the photon spectrum.  The reason is
that these are virtual corrections that yield finite delta-function
contributions at the maximal photon energy, and therefore they contribute
equally to the numerator and the denominator of eq.~(\ref{Moments}).  Thus,
these contributions also cancel to order $\alpha_s$ in the moments $M_n$.

We mentioned earlier that, in the absence of the next-to-leading order result
for the coefficient $C_7^{\,{\rm eff}}$, any calculation of the photon spectrum
(in particular that of \cite{AG}) is renormalization scheme dependent at order
$\alpha_s$.  However, this scheme dependence affects again only the finite part
of the delta-function at the maximal photon energy and therefore drops out from
the moments.

These arguments prove that the moments $M_n$, as calculated below to order
$\alpha_s$, are renormalization scheme and scale independent.

\section{Order $\alpha_s$ QCD corrections}

As we mentioned in the introduction, the heavy quark expansion proves that the
free quark decay model result is the leading term in a systematic expansion in
powers of $1/m_b$, in which the first nonperturbative corrections arise at
order $1/m_b^2$.  In this section we discuss the order $\alpha_s(m_b)$
corrections to the moments of the photon spectrum in the free quark decay
model.  Following \cite{AG}, we keep the strange quark mass finite to
regularize collinear divergencies.  We use the $\overline{\rm MS}$ subtraction
scheme, Feynman gauge, and dimensional regularization for infrared and
ultraviolet divergencies, and phase-space integrals as well.  We verified the
calculation of Ref.~\cite{AG}, and also computed terms that were neglected in
that calculation.  The quantity that is simple to calculate in perturbation
theory is
\begin{equation}\label{moments}
\delta m_n(x_0) = {\displaystyle \int_{x_0}^1 (x^n-1)\,
  {{\rm d}\Gamma_{\scriptscriptstyle\rm FQDM}\over{\rm d}x}\, {\rm d}x \over
  \displaystyle \int_{x_0}^1 {{\rm d}\Gamma_{\scriptscriptstyle\rm FQDM}
  \over{\rm d}x}\, {\rm d}x } \,,
\end{equation}
where $\Gamma_{\scriptscriptstyle\rm FQDM}$ denotes the decay rate in free
quark decay model, and we introduced the dimensionless parameters
\begin{equation}
x = {2E_\gamma\over (1-r)\,m_b} \,,\qquad
r = {m_s^2\over m_b^2}\,.
\end{equation}
The variable $x$ corresponds to $E_\gamma/E_\gamma^{\rm max}$ in the free
quark decay model.  The definition (\ref{moments}) makes it apparent that
the functions $\delta m_n(x_0)$ are proportional to $\alpha_s$ and that they
are not affected by corrections proportional to $\delta(1-x)$ at this order.
In the last section we shall discuss how to relate $\delta m_n$ to the
experimentally measurable moments $M_n$.

The conclusion of Ref.~\cite{AG} is that near the maximal photon energy ($x=1$)
only the operators $O_2$ and $O_7$ are important.  Although we found that the
interference of $O_7$ and $O_8$ is also peaked near $x=1$, this term turns out
to be numerically small \cite{bsg2}.  In this letter we only include the
dominant contributions that were already calculated in \cite{AG}, while we
shall present the full order $\alpha_s$ calculation of the moments $M_n$
elsewhere \cite{bsg2}.

The contributions involving the operator $O_2$ are regular functions of the
photon energy, and their explicit form can be found in \cite{AG}.  The only
singular (and numerically the most significant) contribution to the the free
quark decay rate $\Gamma_{\scriptscriptstyle\rm FQDM}$ at order $\alpha_s$
comes from the operator $O_7$ alone.  In the $r\to0$ limit this contribution
reads (we only explicitly present in this letter the corrections in the $r\to0$
limit, but we included the $r$-dependent terms in our numerical results):
\begin{eqnarray}\label{spect77}
{{\rm d}\Gamma_{77}\over{\rm d}x} &=&
  \Gamma_0\, \bigg[{m_b(\mu)\over m_b}\bigg]^2\,
  \Bigg\{ \bigg[1-{\alpha_s\,C_F\over4\pi}\, \bigg(5 + \frac43\,\pi^2
  - 2\ln{m_b^2\over\mu^2}\bigg)\bigg]\, \delta(1-x) \\*
&+& {\alpha_s\,C_F\over4\pi} \bigg[7+x-2x^2-2(1+x)\ln(1-x)
  - \bigg({7\over1-x}+4\,{\ln(1-x)\over1-x}\bigg)_+\, \bigg] \Bigg\}\,,
  \nonumber
\end{eqnarray}
where $C_F=4/3$ in $SU(3)$, and
\begin{equation}
\Gamma_0 = {G_F^2\, |V_{tb}V_{ts}^*|^2\, \alpha\, C_7^{\,{\rm eff}}(\mu)^2
  \over32\,\pi^4}\, m_b^5 \,.
\end{equation}
By $m_b$ we mean the $b$ quark pole mass at order $\alpha_s$, as discussed
below.  The $[f(x)]_+$ distribution corresponding to a function $f(x)$ acts
on a test function $g(x)$ as
\begin{equation}
\int_0^1 [f(x)]_+\,g(x)\,{\rm d}x = \int_0^1 f(x)\,[g(x)-g(1)]\,{\rm d}x\,.
\end{equation}
By including the $\mu$-dependence in eq.~(\ref{spect77}), we can explicitly
verify that $\Gamma_{77}$ is $\mu$-independent to order $\alpha_s$.  This
cancellation by itself does not reduce significantly the $\mu$-dependence of
the theoretical prediction for the total decay rate, as the dominant part of
$C_7^{\,{\rm eff}}(\mu)$ arises from the mixing of $O_2$ with $O_7$ at order
$\alpha_s$, but still at leading log.  The $\alpha_s$ correction in
eq.~(\ref{spect77}) modifies the prediction for the total decay rate by about
15\%, while it affects the first two moments of the photon spectrum, $M_1$ and
$M_2$, by less than 3\% and 5\%, respectively.  Our numerical results for
$\delta m_1$ and $\delta m_2$, including the contribution of $O_2$, are shown
in Table~I.

When evaluating these corrections, it has to be kept in mind that the
perturbative expansion becomes singular in the photon endpoint region, and a
resummation of the perturbative corrections may be required.
As we are interested in the moments $M_n$ for small $n$, and
$(\alpha_sC_F/2\pi)\ln^2(1-x)$, the exponent of the Sudakov factor that
suppresses the endpoint spectrum \cite{exp}, becomes of order unity only around
$x\sim0.99$, our calculation is consistent without taking these effects into
account.

\section{Nonperturbative corrections}

We include the nonperturbative corrections to the free quark decay to order
$1/m_b^2$ and leading order in $\alpha_s$.  At that order only corrections to
the matrix element of the operator $O_7$ contribute.  We present here the
resulting corrections to the photon spectrum \cite{FLS} only in the $r\to0$
limit again:
\begin{equation}\label{npspec}
{{\rm d}\Gamma\over{\rm d}x} = \Gamma_0\, \bigg( 1
  + {\lambda_1-9\lambda_2\over2m_b^2} \bigg)\, \bigg[  \delta(1-x) -
  {\lambda_1+3\lambda_2\over2m_b^2}\, \delta'(1-x) - {\lambda_1\over6m_b^2}\,
  \delta''(1-x) \bigg] \,.
\end{equation}
The dimensionful constants $\lambda_1$ and $\lambda_2$ parameterize the matrix
elements of the kinetic and chromomagnetic operators, respectively, which
appear in the Lagrangian of the HQET at order $1/m_Q$:
\begin{eqnarray}\label{ldef}
\lambda_1 &=& {1\over2}\,
  \langle M(v)|\,\bar h_v\,(iD)^2\,h_v\,|M(v)\rangle\,, \nonumber\\*
\lambda_2 &=& {1\over2d_M}\, \langle M(v)|\,{g_s\over2}\,\bar h_v\,
  \sigma_{\alpha\beta}\,G^{\alpha\beta}\,h_v\,|M(v)\rangle\,,
\end{eqnarray}
where $d_P=3$ and $d_V=-1$ for pseudoscalar and vector mesons, respectively.
$M(v)$ denotes the meson state and $h_v$ denotes the quark field of the
effective theory with velocity $v$.  The numerical value of $\lambda_2$ can be
extracted from the mass splitting between the vector and pseudoscalar mesons,
$\lambda_2=(m_{B^*}^2-m_B^2)/4\simeq0.12\,{\rm GeV}^2$, while there is no
similarly simple way to determine $\lambda_1$ from experiments (see,
{\it e.g.}, \cite{YZ}).  Without including the order $\alpha_s$ corrections
discussed in the previous section, one obtains to order $1/m_b^2$ in the heavy
quark expansion
\begin{eqnarray}\label{basic}
\langle E_\gamma \rangle &=& {m_b\over2}\,
  \bigg(1-{\lambda_1+3\lambda_2\over2m_b^2}\bigg) = {M_B-\bar\Lambda\over2} \,,
  \nonumber\\*
\langle E_\gamma^2\, \rangle - \langle E_\gamma \rangle^2 &=&
  -{\lambda_1\over12} \,.  \end{eqnarray}
In general, the central moments
$\langle (E_\gamma - \langle E_\gamma \rangle)^n \rangle$ for $n\geq2$ are
proportional to $\lambda_1(m_b/2)^{n-2}$.  Therefore, they are particularly
useful for measuring $\lambda_1$.  This is not unexpected, since $\lambda_1$
is the measure of the Fermi motion of the $b$ quark that is responsible for
the smearing of the photon spectrum.  A comparison of the values of
$\lambda_1$ extracted from different central moments can be used to estimate
the systematic errors.

The parameter $\bar\Lambda$ in eq.~(\ref{basic}) describes the mass difference
between a heavy meson and the heavy quark that it contains, and it is one of
the parameters that set the scale of the $1/m$ expansion \cite{FNL}.  It is
related to the $B$ meson mass and the $b$ quark pole mass according to
\begin{equation}\label{Ldef}
M_B = m_b + \bar\Lambda - {\lambda_1+3\lambda_2\over2m_b} \,.
\end{equation}
The quantity $\bar\Lambda$ suffers from renormalon ambiguities \cite{renorm}.
However, at any finite order in $\alpha_s$ it is consistent to extract
$\bar\Lambda$ (or the pole mass $m_b$) from certain experiment(s), and use the
resulting numerical values to evaluate theoretical predictions accurate to the
same order in perturbation theory for other processes \cite{rencan}.

The series of the nonperturbative corrections to the moments of the photon
spectrum, $M_n(E_0)$, is under control if the invariant mass of the final
hadronic state corresponding to the lower cut $E_0$ is above the resonance
region.  Eq.~(\ref{npspec}) is only related to the experimentally measured
spectrum once the theoretical expression is smeared over typical hadronic
scales.  This smearing is provided by taking the moments $M_n(E_0)$, if $n$
is not too large, and $E_0$ is sufficiently far from $E_\gamma^{\rm max}$.
Using the relation
\begin{equation}
{M_B^2 - M_{X_s}^2\over 2M_B} = E_\gamma \,,
\end{equation}
we see that the present experimental signal region of $E_\gamma>2.2\,$GeV
\cite{CLEO} corresponds to $M_{X_s}\lesssim2.2\,$GeV.  Even below this scale
the widths of the $X_s$ resonances are typically larger than their mass
differences, and there are no resonances above $2.5\,$GeV \cite{PDG}.  It is
still important for the reliability of our analysis to try to lower the
experimental cut on the photon energy.  For example, $E_0=2\,$GeV or
$E_0=1.8\,$GeV would correspond to $M_{X_s}\lesssim2.6\,$GeV or
$M_{X_s}\lesssim3\,$GeV, respectively.  Varying $E_0$ provides a check on the
systematic uncertainties: the extracted values of $\bar\Lambda$ and $\lambda_1$
should be unaffected by the variations of $E_0$, once the corresponding
hadronic invariant mass is sufficiently above the resonance region.

\section{Summary and conclusions}

We can summarize our discussion by writing the theoretical prediction for
the first two moments of the photon spectrum as
\begin{eqnarray}\label{final}
M_1(E_0) &=& {M_B-\bar\Lambda\over2}\, \bigg[ 1 +
  \delta m_1 \bigg({2E_0\over m_b}\bigg) + {\cal O}\bigg(\alpha_s^2,\,
  \alpha_s\,{\Lambda^2\over m_b^2},\,
  {\Lambda^3\over m_b^3} \bigg)\bigg] \,,\\*[4pt]
M_2(E_0) - M_1(E_0)^2 &=& -{\lambda_1\over12} + \bigg({m_b\over2}\bigg)^{\!2}\,
  \bigg[ \delta m_2 \bigg({2E_0\over m_b}\bigg)
  - 2\,\delta m_1 \bigg({2E_0\over m_b}\bigg) + {\cal O}\bigg(\alpha_s^2,\,
  \alpha_s\,{\Lambda^2\over m_b^2},\, {\Lambda^3\over m_b^3} \bigg)\bigg] \,.
  \nonumber
\end{eqnarray}
We used $\Lambda$ to denote some QCD scale of order $\Lambda_{\rm QCD}$ or
$\bar\Lambda$.  To the order these relations are accurate, it is consistent to
replace $m_b$ by $M_B-\bar\Lambda$ everywhere in eq.~(\ref{final}).
The numerical results for $\delta m_1(x_0)$ and $\delta m_2(x_0)$ are
listed in Table~I for three different values of $x_0$.

The significance of these relations is that they provide a reliable means
of determining $\bar\Lambda$ (that is, $\bar\Lambda$ at order $\alpha_s$
\cite{rencan}) and $\lambda_1$, or equivalently, measure the $b$ quark pole
mass at order $\alpha_s$.  Especially the first relation in eq.~(\ref{final})
is remarkable, since it is independent of $\lambda_1$ (and $\lambda_2$), and
very sensitive to $\bar\Lambda$, with small theoretical uncertainties.
The sensitivity of the second relation to $\lambda_1$ is reduced because of
the factor 1/12.

We would like to emphasize that the left-hand side of the relations in
eq.~(\ref{final}) are measurable at CLEO; in fact, the central
values can be extracted from Ref.~\cite{CLEO}.  As we do not know the
cross-correlations of the errors on the data points, we are not in a
position to quote numerical values for the experimental uncertainties.
{}From the central values of the data, solving the first equation in
(\ref{final}), we find (with large uncertainties)
\begin{equation}
\bar\Lambda \sim 450 \,{\rm MeV}\,, \qquad  m_b\sim 4.83 \,{\rm GeV}\,.
\end{equation}
We do not quote even a central value for $\lambda_1$, as the present
experimental data do not constrain it to any reasonable accuracy.
By varying all input parameters ($C_7(\mu)$ and $C_2(\mu)$ corresponding to the
range $m_b/2<\mu<2m_b$, $\alpha_s$ corresponding to $0.11<\alpha_s(M_Z)<0.13$,
$m_s$ between $100\,$MeV and $500\,$MeV, $m_t$, and $m_c$) we find that the
theoretical uncertainty of this measurement of $\bar\Lambda$ will be as small
as $\pm30\,$MeV, while that of $\lambda_1$ about $\pm0.15\,{\rm GeV}^2$.

In view of our earlier discussion, it is important to try to expand the
experimental signal region.  On the one hand, the systematic uncertainties
inherent in our analysis (related to how well duality holds) can be
estimated by varying the lower cut on the photon energy, as discussed at the
end of Section V.  On the other hand, expanding the signal region would
diminish the sensitivity of the results as to whether the Sudakov logarithms
at the endpoint are resummed or not.

The sensitivity of the moments of the spectrum to new physics is limited by how
much operators other than $O_7$ affect $M_n$.  We found that $O_2$ does not
contribute to $\delta m_1$ and $\delta m_2$ by more than 10\% in the SM.  Given
that the experimental constraint on the total decay rate from CLEO excludes
large deviations from the SM, we conclude that the moments are largely
insensitive to new physics.  Even if physics beyond the SM contributes to the
$B\to X_s\,\gamma$ decay, the proposed determination of $\bar\Lambda$ and
$\lambda_1$ is likely to remain unaffected.

We conclude that the moments of the photon spectrum in the inclusive
$B\to X_s\,\gamma$ decay will provide reliable measurements of fundamental
parameters of QCD, which in turn will refine theoretical predictions for other
observables in heavy quark decays.  The complete order $\alpha_s$ calculation
of the moments and a more detailed analysis of the theoretical uncertainties
will be presented in a forthcoming paper \cite{bsg2}.

\acknowledgements
We are grateful to Peter Cho, David Politzer, Ira Rothstein, Alan Weinstein
and especially Mark Wise for numerous discussions.

{\tighten

} 

\begin{table}
   \begin{tabular}{cl|ccc}
& &			$x_0=0.91$ & $x_0=0.83$ & $x_0=0.75$ \\
\hline
$\delta m_1(x_0)$ & $r=4\cdot10^{-4}$~~ & $-0.014$ & $-0.020$ & $-0.025$  \\
& $r=1\cdot10^{-2}$ & 			  $-0.012$ & $-0.017$ & $-0.020$  \\
\hline
$\delta m_2(x_0)$ & $r=4\cdot10^{-4}$~~ & $-0.028$ & $-0.040$ & $-0.046$ \\
& $r=1\cdot10^{-2}$ & 			  $-0.023$ & $-0.032$ & $-0.038$
   \end{tabular}
\caption{%
Central values of $\delta m_1(x_0)$ and $\delta m_2(x_0)$ for two different
values of $m_s$.  $r=4\cdot10^{-4}$ corresponds to $m_s=100\,$MeV, while
$r=1\cdot10^{-2}$ corresponds to a constituent quark mass $m_s=500\,$MeV.
For $m_b\simeq4.8\,$GeV, $x_0=0.91$, $x_0=0.83$ and $x_0=0.75$ correspond to
$E_0=2.2\,$GeV, $E_0=2.0\,$GeV and $E_0=1.8\,$GeV, respectively.}
\end{table}

\end{document}